\documentclass[10pt]{article}

\usepackage{amsmath}
\usepackage{amssymb}
\usepackage{graphics}
\usepackage{rotating}
\usepackage{cite}
\usepackage{color}
\usepackage{fancybox}
\usepackage{pstricks}

\def\0{\mbox{\tiny $0$}}
\def\1{\mbox{\tiny $1$}}
\def\2{\mbox{\tiny $2$}}
\def\3{\mbox{\tiny $3$}}
\def\4{\mbox{\tiny $4$}}
\def\5{\mbox{\tiny $5$}}
\def\6{\mbox{\tiny $6$}}
\def\7{\mbox{\tiny $7$}}
\def\8{\mbox{\tiny $8$}}
\def\9{\mbox{\tiny $9$}}
\def\B{\mbox{\tiny $\Box$}}
\def\N{\mbox{\tiny $\triangle$}}
\title{\shadowbox{\large \bf LIGHT TRANSMISSION THROUGH A TRIANGULAR AIR GAP}}

\author{
\small  Silv\^ania A. Carvalho\,\,\,\,and\,\,\,\,Stefano De
Leo\thanks{deleo@ime.unicamp.br} \\
\small Department of Applied Mathematics, State University of
Campinas, Brazil}

\date{\small
\fcolorbox{black}{yellow} {\color{red} $\bullet$ {\color{black}{
{\footnotesize   {\sc Journal of Modern Optics} {\bf 60}, 437-443 (2013)
}}} {\color{red}{$\bullet$}} } }

\begin{document}
%

\maketitle

\vspace*{-.7cm}

\begin{abstract}
\noindent Due to the recent interest in studying propagation of light through triangular
air gaps,  we calculate, by using the analogy between optics and quantum mechanics and the multiple step technique, the transmissivity through a triangular air gap surrounded by an homogeneous
dielectric medium. The new formula is then compared with the formula used in literature. Starting from the qualitative and quantitative differences between these formulas, we propose optical experiments to test our theoretical results.
\end{abstract}












\section*{\normalsize I. INTRODUCTION}

Propagation of light through  rectangular air gaps situated between homogeneous media is
surely one of the most intriguing phenomenon of optics\cite{op1,op2,op3}. Although studied for hundred of years\cite{op4,op5,op6}, it still represents matter of discussion and challenge for theoretical and experimental investigations\cite{te1,te2,te3,te3bis,te4,te5,te6,te7,te8}.
 For a  plane, time-harmonic, electromagnetic wave moving along the $z$-axis, the transmissivity through a rectangular air gap, surrounded by  an homogeneous dielectric
whose properties are constant throughout each plane perpendicular to a fixed direction   $z_*$,
is given by\cite{te6,eq11,eq12}
\begin{equation}
\label{tb}
T_{\B}^{^{(s,\,p)}}=\left[\,1 +\, \alpha_{\B}^{^{(s,\,p)}}\,
\sin^{\2}\left( \sqrt{1-n^{^2}\sin^{\2}\theta_*}\,\,\,kL_{*} \right)\right]^{^{-1}}\,\,.
\end{equation}
The upper index $s$ and $p$ respectively refer to waves linearly  polarized with its electric and magnetic vector perpendicular to the $y$-$z$ incidence plane, the so-called transverse electric (TE) and magnetic (TM) waves\cite{op1}. The incidence angle  $\theta_*$ is the angle between the beam motion ($z$ axis) and the stratification direction
($z_*$ axis), $k=2\pi/\lambda$ is the wave number, $L_*$ is the distance between the dielectric blocks of refractive index $n$,
 \[ \alpha_{\B}^{^{(s)}}=(n^{\2}-1)^{^{2}}
\,\mbox{\Large /}\,\left[\,4\,n^{\2} (1-n^{\2}\sin^{\2}\theta_*)\cos^{\2}\theta_{*}\, \right]
\,\,\,\,\,\,\,\mbox{and}\,\,\,\,\,\,\,
 \alpha_{\B}^{^{(p)}} = \alpha_{\B}^{^{(s)}} \left[(n^{\2}+1)\sin^{\2}\theta_{*} - 1 \right]^{^{2}} \,\,.
\]
The geometric layout is drawn in Fig.\,1a. For an incidence angle which tends to the critical angle $\theta_{*}^{^{c}}=\arcsin(1/\,n)$, the transmissivity is given by
\begin{equation}
\label{tb2}
T_{\B}[\theta_*\to\theta_*^{^{c}}]\,\,\,\to\,\,\,\left\{\begin{array}{ll}
\left[\,1 + (n^{\2}-1)\,\pi^{\2}\,(L_*/\lambda)^{^{2}}\right]^{^{-1}} & \mbox{$s$-waves}\,\,,\\
\left[\,1 + (n^{\2}-1)\,(\pi/n^{\2})^{^2}\,(L_*/\lambda)^{^{2}}\right]^{^{-1}} & \mbox{$p$-waves}\,\,.\end{array}
\right.
\end{equation}
In this case, for $L_*\gg \lambda$  no light is transmitted in the second dielectric block. Decreasing the value of $L_*$ up to several wavelengths, waves pass through the air interspace and light is transmitted in the second dielectric block.   Due to the fact that varying the thickness of an air gap with precision is a very hard task, it has been proposed to study the transmission through a {\em triangular air gap}\cite{tri}. This allows, by changing the incidence  point of the light on the first dielectric/air interface, $h$,   to study the propagation through air gaps of different thickness, see Fig.\,1b. Observing that the transmission probability  is  highly sensitive to the air gap between the dielectric blocks, coherence phenomena  can be only seen for very small angles, i.e. $\varphi\ll 1$. The distance  between the dielectric blocks, $L_*$,  can be then approximated by $h\,\varphi$ and we can use for  the transmissivity through  a triangular air gap the following formula\cite{te6},
\begin{equation}
\label{tb3}
T_{\B \to \N }^{^{(s,\,p)}}:=T_{\B}^{^{(s,\,p)}}[\,L_*\to h\,\varphi\,]\,\,.
\end{equation}
For $n^{\2}\sin^{\2}\theta_*<1$, the incident beam is in general divided between two secondary beams.  Nevertheless, there is a situation in which we have only the transmitted beam. The phenomenon of {\em total  transmission} happens when $h$ is an integer multiple of $\lambda\,/2\,\varphi\,\sqrt{1-n^{^{2}}\sin^{\2}\theta_*}$.
For $n^{\2}\sin^{\2}\theta_*\to 1$, the triangular air gap acts as a frequency filter and only wavelengths of the order of $h\,\varphi$ are transmitted from the first to the second dielectric block.

The formulas used for the transmission of light through three layers, see Eq.\,(\ref{tb}), are classical formulas. It is highly desirable to find a quantum analog.
Starting from the analogy between optics\cite{op1,op2,op3} and quantum mechanics\cite{qm1,qm2} and by using the multiple step technique\cite{ms1,ms2,ms3}, we calculate  the propagation of an electromagnetic wave through a {\em triangular} air gap.  Then, for $\varphi\ll 1$, we analyze the differences between the {\em new} formula and the approximated one given in Eq.\,(\ref{tb3}) and used in litterature\cite{te6}.

The paper is structured as follows. In section II, we discuss the analogies between light interaction upon the  dielectric/air interface and the step potential in non relativistic quantum mechanics (NRQM).  In section III, for the convenience of the reader, we briefly review the multiple step calculation\cite{ms1,ms2,ms3,Anderson}  which leads, for a rectangular air gap, to the transmission amplitude given in  Eq.\,(\ref{tb}). In section IV, we consider  $\varphi$-rotation of the second interface in the multiple step calculation and give the {\em new} transmission formula. In the final section, we draw our conclusions. In particular, we shall compare our formula with the formula used in literature, Eq.\,(\ref{tb3}),  and propose  optical experiments for testing the {\em new} formula.

\section*{\normalsize II. ANALOGY BETWEEN OPTICS AND QUANTUM MECHANICS}

In this section, we  present the connection between optical and quantum mechanics problems.
In particular, we show how the NRQM formalism can be easily used to obtain the well-known  Fresnel formulas for reflection and transmission of $s$ and $p$ polarized light. This connection will be fundamental for the calculation of the transmission probabilities done in the following sections.

Let us consider the plane $x=0$ as the incidence plane. The incident ray  travels in the first dielectric block of refractive index $n$ along the $z$-axis,
\[ \boldsymbol{q}=n k\,\boldsymbol{\hat{z}}\,\,.\]
In terms of the new $y_*$-$z_*$ axes, the incident wave number can be rewritten as follows
\begin{equation}
\left(\begin{array}{c} q_{y_*}\\ q_{z_*} \end{array} \right) =
\left(
\begin{array}{rr} \cos \theta_{*} & \sin \theta_{*} \\
-\,\sin \theta_{*} & \cos \theta_{*}
\end{array}   \right)\left(\begin{array}{c} q_y\\ q_z \end{array}
\right) = nk\,\,\left(\begin{array}{c} \sin \theta_{*} \\ \cos \theta_{*}
\end{array} \right)\,\,.
\end{equation}
 The dynamics of  TE  waves, which moving in the first dielectric block  encounter the  dielectric/air discontinuity $d_*$, is governed by\cite{op1}
 \begin{equation}
 \frac{\partial^{\2}\,\,}{\partial y_*^{^{2}}}\,\Psi^{^{(s)}} + \frac{\partial^{\2}\,\,}{\partial z_*^{^{2}}}\,\Psi^{^{(s)}} + k^{\2}n^{\2}(z_*)\,\Psi^{^{(s)}}=0\,\,\,,\,\,\,\,\,\,\,\,\,\,\,\,\mbox{with}\,\,\,\,\,
 n(z_*)=\left\{\begin{array}{lcl} n &\,\,\,\mbox{for} & z_*<d_*\\
 1 &\,\,\,\mbox{for} & z_*>d_* \end{array}\right.\,\,.
 \end{equation}
 This equation,  by taking  the correspondence
 \[
 n(z_*)\,\,\,\to\,\,\,\sqrt{2m\,[E-V(z_*)]}\,/\,k\hbar\,\,\,,\,\,\,\,\,\,\,\,\,\,\,\,
 V(z_*)=\left\{\begin{array}{lcl} V_{\0} &\,\,\,\mbox{for} & z_*<d_*\\
 0 &\,\,\,\mbox{for} & z_*>d_* \end{array}\right.\,\,,\]
 exactly mimics the NRQM step  potential dynamics\cite{qm1}. This is obviously a mathematical analogy. The Maxwell and Schr\"odinger equations describe a different physics and such equations only  coincide, in their mathematical description, for special cases. This correspondence is, for example, not exact for TM waves. As it will be shown later,  for TM  waves we obtain the Fresnel formulas by introducing a translation rule. The TE incident and reflected beams which move in the first dielectric block, $z_*<d_*$,  are represented by the wave function
 \[\Psi_{_{<}}^{^{(s)}}=\left\{\,\exp[\,i\,q_{z_*}z_*\,]\, +\,\, r^{^{(s)}}\,\exp[\,-\,i\,q_{z_*}z_*\,]\,\right\}\,\exp[\,i\,q_{y_*}y_*\,] \,\,,\]
 where $q_{y_*}$ and $q_{z_*}$ are the wave number components of the beam which propagates in the first dielectric block. The transmitted beam which moves in the air gap, $z_*>d_*$, is represented by
 \[\Psi_{_{>}}^{^{(s)}}= t^{^{(s)}}\,\exp[\,i\,p_{z_*}z_*\,]\,\exp[\,i\,p_{y_*}y_*\,] \,\,,\]
$p_{y_*}$ and $p_{z_*}$ are now the wave number components of the beam which propagates in the air gap.  Being the discontinuity along the $z_*$ axis, the $y_*$ component of the wave number does not change passing from the dielectric to air, i.e. $p_{y_*}=\,q_{y_*}=nk\,\sin\theta_*$. This implies
\begin{equation}
p_{z_*} = k\,
\sqrt{1 - n^{\2} \sin^{\2}\theta_{*}}\,\,.
\end{equation}
From the matching conditions, $\Psi_{_{<}}(z_*=d_*)=\Psi_{_{>}}(z_*=d_*)$ and  $\Psi^{^{\prime}}_{_{<}}(z_*=d_*)=\Psi^{^{\prime}}_{_{>}}(z_*=d_*)$, we find\cite{VF}
\begin{eqnarray}
r^{^{(s)}}[q_{z_*},p_{z_*},d_*] &=& \frac{q_{z_{*}}-\,p_{z_*}}{q_{z_*}+\,
p_{z_*}}\,\,\,\exp[\,2\,i\,q_{z_*}d_*\,]\,\,, \nonumber \\
t^{^{(s)}}[q_{z_*},p_{z_*},d_*] & = & \frac{2\,q_{z_{*}}}{q_{z_*}+\,
p_{z_*}}\,\,\,\exp[\,i\,(q_{z_*}-\,p_{z_*})\,d_*\,]\,\,.
\end{eqnarray}
The Fresnel formulas are immediately obtained from the previous amplitudes,
\begin{eqnarray}
R^{^{(s)}} &=&|\,r^{^{(s)}}|^{^{2}}= \left(\frac{n \cos\theta_{*} - \sqrt{1 - n^{\2} \sin^{\2}\theta_{*}}  }{n \cos\theta_{*} +  \sqrt{1 - n^{\2} \sin^{\2}\theta_{*}}  }\, \right)^{^{2}} \,\,,\nonumber\\
T^{^{(s)}} & = & \frac{p_{z_*}}{q_{z_*}}\,|\,t^{^{(s)}}|^{^{2}}=  \frac{ 4 \,n \cos\theta_{*} \sqrt{1 - n^{\2} \sin^{\2}\theta_{*}}  }{(n \cos\theta_{*} + \sqrt{1 - n^{\2} \sin^{\2}\theta_{*}})^{^{2}}}\,\,.
\end{eqnarray}
For $n\,\sin\theta_*\to 1$, $p_{z_*}$ tends to zero and we have total reflection, $|r^{^{(s)}}|=1$.  The reflection and transmission amplitudes for TM waves can be obtained by using the following translation rules,
\begin{eqnarray}
r^{^{(p)}}[q_{z_*},p_{z_*},d_*] &=&  r^{^{(s)}}\left[\frac{q_{z_*}}{n},np_{z_*},0\right]\,\,\,e^{2\,i\,q_{z_*}d_*}\,\,, \nonumber \\
t^{^{(p)}}[q_{z_*},p_{z_*},d_*] & = & t^{^{(s)}}\left[\frac{q_{z_*}}{n},np_{z_*},0\right]\,\,\,\,\,\,e^{i\,(q_{z_*}-\,p_{z_*})\,d_*}\,\,.
\end{eqnarray}
For $n\,\sin\theta_*<1$, the incident beam is, in general, partially reflected and transmitted with intensities
\begin{eqnarray}
R^{^{(p)}} &=&|\,r^{^{(p)}}|^{^{2}}= \left(\frac{\cos\theta_{*} - n\,\sqrt{1 - n^{\2} \sin^{\2}\theta_{*}}  }{\cos\theta_{*} + n\, \sqrt{1 - n^{\2} \sin^{\2}\theta_{*}}  }\, \right)^{^{2}} \,\,,\nonumber\\
T^{^{(p)}} & = & n^{\2}\,\frac{p_{z_*}}{q_{z_*}}\,|\,t^{^{(p)}}|^{^{2}}=  \frac{ 4 \,n \cos\theta_{*} \sqrt{1 - n^{\2} \sin^{\2}\theta_{*}}  }{(\cos\theta_{*} + n\,\sqrt{1 - n^{\2} \sin^{\2}\theta_{*}})^{^{2}}}\,\,.
\end{eqnarray}
For a particular angle,
\[ \sin\theta^{^{B}}_*=1\,/\,\sqrt{n^{^{2}}+1}\,\,,\]
known as Brewster´s angle\cite{op1,op2,op3}, we have $q_{z_*}=n^{\2}p_{z_*}$ which implies  $r^{^{(p)}}=0$ and, consequently, the incident light is totally  transmitted.

\section*{\normalsize III. MULTIPLE STEP ANALYSIS FOR A RECTANGULAR AIR GAP}

It was demonstrated in previous works\cite{te7,te8} that, the transmission and reflection amplitudes of a rectangular (air) interspace can be built up by successive application of the step analysis\cite{Anderson,ms1,ms2,ms3}. As illustrated in Fig.\,2, at the first dielectric/air surface, $z_*=0$, the incident wave is divided in two plane waves.  The transmitted wave with amplitude $t^{^{(s,p)}}[q_{z_*},p_{z_*},0]$ moves through the air gap and after touching the second air/dielectric surface, $z_*=L_*$, will be transmitted with amplitude
\[ t^{^{(s,p)}}[q_{z_*},p_{z_*},0]\,\,t^{^{(s,p)}}[p_{z_*},q_{z_*},L_*]\,\,.\]
This  represents the first contribution to transmission through the rectangular air gap. The complete transmission amplitude is built up by  considering the loop factor obtained by multiplying the amplitude of the reflected wave at the second air/dielectric surface  by the one
of the reflected wave at the first air/dielectric surface,
\[  r^{^{(s,p)}}[p_{z_*},q_{z_*},L_*]\,\,r^{^{(s,p)}}[-p_{z_*},-q_{z_*},0]\,\,.\]
If overlaps dominate, a single beam is transmitted and to find the transmission amplitude we have to sum all coherent multiple contributions,
    \begin{eqnarray}
    \label{amp}
t_{\B}^{^{(s,p)}} & = & t^{^{(s,p)}}[q_{z_*},p_{z_*},0]\,t^{^{(s,p)}}[p_{z_*},q_{z_*},L_{*}] \sum ^{\infty}_{m=0} \left(r^{^{(s,p)}}[p_{z_*},q_{z_*},L_{*}]\, r^{^{(s,p)}}[-p_{z_*},-q_{z_*},0]\right)^{^m} \nonumber \\
\hspace{2cm}& = & \frac{t^{^{(s,p)}}\left[q_{z_*},p_{z_*},0\right] t^{^{(s,p)}}\left[p_{z_*},q_{z_*},L_*\right]}{ 1- r^{^{(s,p)}}\left[p_{z_*},q_{z_*},L_*\right] r^{^{(s,p)}}\left[-p_{z_*},-q_{z_*},0\right]}\,\,.
\end{eqnarray}
Let us first calculate  the transmissivity for the case of  TE waves. Observing that
\begin{eqnarray}
t_{\B}^{^{(s)}} &=&    4\,q_{z_*}p_{z_*}\,e^{-\, i\, q_{z_*} L_{*}}\,  \left[\,\left(q_{z_*}+\,p_{z_*}\right)^{^{2}}e^{-\, i\, p_{z_*} L_{*}} - \left(q_{z_*}-\,p_{z_*}\right)^{^{2}} e^{i\, p_{z_*} L_{*}}   \right]^{^{-1}} \nonumber\\
  & = & e^{- i\, q_{z_*} L_{*}}\,\left[ \cos(p_{z_*} L_{*}) - i\, \frac{q^{\2}_{z_*} + p^{\2}_{z_*}}{2\,q_{z_*}\,p_{z_*}}\, \sin(p_{z_*} L_{*}) \right]^{^{-1}}\,\,,
\end{eqnarray}
we have
\begin{equation}
\label{tbs}
T_{\B}^{^{(s)}}=  \left|t_{\B}^{^{(s)}}\right|^{^{2}} =\left[1 + \left(\frac{q_{z_*}^{\2} - p_{z_*}^{\2}}{2\, q_{z_*} p_{z_*}}\right)^{^{2}} \sin^{\2}\left(p_{z_*} L_{*}\right)\right]^{^{-1}}\,\,.
\end{equation}
A simple calculation shows that
\begin{equation}
\frac{(q_{z_*}^{\2} - p_{z_*}^{\2})^{^2}}{4\, q_{z_*}^{^{2}} p_{z_*}^{^{2}}} =
\frac{(n^{\2}-1)^{^{2}}}{\,4\,n^{\2} \cos^{\2}\theta_{*}
(1-n^{\2}\sin^{\2}\theta_*)} =  \alpha_{\B}^{^{(s)}}\,\,,
\end{equation}
and, consequently, we reproduce Eq.\,(\ref{tb}) for $s$-polarized waves. In a similar way, observing that
\begin{equation}
T_{\B}^{^{(p)}}=\left[1 + \left(\frac{q_{z_*}^{\2}/n^{\2} - n^{\2}\,p_{z_*}^{\2}}{2\, q_{z_*} p_{z_*}}\right)^{^{2}} \sin^{\2}\left(p_{z_*} L_{*}\right)\right]^{^{-1}}\,\,,
\end{equation}
and
\begin{equation}
\frac{( q_{z_*}^{\2} / n^{\2} - n^{\2} p_{z_*}^{\2} )^{^2}}{4\, q_{z_*}^{^2} p_{z_*}^{^2}} =
\frac{(\cos^{\2}\theta_{*}-n^{\2}+ n^{\4} \sin^{\2}\theta_{*})^{^{2}}}{4\, n^{\2}  \cos^{\2}\theta_{*} (1-n^{\2}\sin^{\2}\theta_*)} = \frac{(n^{\2}-1)^{^{2}} \left[(n^{\2}+1)\sin^{\2}\theta_{*} - 1 \right]^{^{2}}}{\,4\,n^{\2} \cos^{\2}\theta_{*}
(1-n^{\2}\sin^{\2}\theta_*)} =  \alpha_{\B}^{^{(p)}}\,\,,
\end{equation}
we obtain  Eq.\,(\ref{tb}) for TM waves.

\section*{\normalsize IV. TRANSMISSION THROUGH A TRIANGULAR AIR GAP}

For the first dielectric/air interface, the computation of the transmissivity through the triangular air gap follows  the calculation done in the previous section. For the second discontinuity, we have to observe that the second dielectric block is now rotated, with respect to the original layout, by an angle $\varphi$
\[
\left(\begin{array}{c} \widetilde{y}\\ \widetilde{z} \end{array}
\right) = \left(
\begin{array}{rr} \cos \varphi & - \sin \varphi \\
\,\sin \varphi & \cos \varphi
\end{array}   \right)\left(\begin{array}{c} y_*\\ z_* \end{array}
\right)\,\,.
\]
The  transmission through a triangular air gap can be immediately obtained from Eq.(\ref{amp}),
by substituting the first transmitted wave amplitude and the loop factor respectively by
\[t^{^{(s,p)}}[q_{z_*},p_{z_*},0]\,\, t^{^{(s,p)}}[p_{\widetilde{z}},q_{\widetilde{z}},\widetilde{L}]\,\,\,\,\,\,\,\mbox{and}\,\,\,\,\,\,\,
 r^{^{(s,p)}}[p_{\widetilde{z}},q_{\widetilde{z}},\widetilde{L}]\,\,r^{^{(s,p)}}[-p_{z_*},-q_{z_*},0]
 \,\,,\]
with
\[
\left(\begin{array}{c} p_{\widetilde{y}}\\ p_{\widetilde{z}} \end{array}
\right) = \left(
\begin{array}{rr} \cos \varphi & - \sin \varphi \\
\,\sin \varphi & \cos \varphi
\end{array}   \right)\left(\begin{array}{c} p_{y_*}\\ p_{z_*} \end{array}
\right)\,\,\,\,\,\,\,\mbox{and}\,\,\,\,\,\{\,q_{\widetilde{y}}\,,\,q_{\widetilde{z}}\,\}=\left\{\, p_{\widetilde{y}}\,,\,\sqrt{n^{\2}k^{^{2}} - q^{\2}_{\widetilde{y}}}\,\right\}\,\,.
\]
We then find
 \begin{equation}
    \label{amp2}
t_{\N}^{^{(s,p)}}  = \frac{t^{^{(s,p)}}\left[q_{z_*},p_{z_*},0\right] t^{^{(s,p)}}[p_{\widetilde{z}},q_{\widetilde{z}},\widetilde{L}]}{ 1- r^{^{(s,p)}}[ p_{\widetilde{z}},q_{\widetilde{z}},\widetilde{L}] r^{^{(s,p)}}\left[-p_{z_*},-q_{z_*},0\right]}\,\,,
\end{equation}
Let us first discuss the case of TE waves and rewrite Eq.\,(\ref{amp2}) as follows
\begin{eqnarray}
t_{\N}^{^{(s)}} &=&    4\,q_{z_*}p_{\widetilde{z}}\,e^{-\, i\, q_{\widetilde{z}} \widetilde{L}}\,  \left[\,\left(q_{z_*}+\,p_{z_*}\right)\left(q_{\widetilde{z}}+\,p_{\widetilde{z}}\right)\,e^{-\, i\, p_{\widetilde{z}} \widetilde{L}} - \left(q_{z_*}-\,p_{z_*}\right)\left(q_{\widetilde{z}}-\,p_{\widetilde{z}}\right)\, e^{i\, p_{\widetilde{z}} \widetilde{L}}   \right]^{^{-1}} \nonumber\\
  & = & e^{-\, i\, q_{\widetilde{z}} \widetilde{L}}\,\left[ \frac{ q_{z_*}p_{\widetilde{z}}+ p_{z_*}q_{\widetilde{z}}}{2\,q_{z_*}p_{\widetilde{z}}}\, \cos(p_{\widetilde{z}} \widetilde{L}) - i\, \frac{q_{z_*}q_{\widetilde{z}} + p_{z_*}p_{\widetilde{z}} }{2\,q_{z_*}\,p_{\widetilde{z}}}\, \sin(p_{\widetilde{z}} \widetilde{L}) \right]^{^{-1}}\,\,.
\end{eqnarray}
Observing that at the first discontinuity, $z_*=0$, the transmissivity is obtained by multiplying  $|t^{^{(s)}}\left[q_{z_*},p_{z_*},0\right]|^{^2}$ by the factor  $p_{z_{*}}/q_{z_{*}}$, whereas at the second discontinuity, $\widetilde{z}=\widetilde{L}$, the transmissivity is obtained by multiplying    $|t^{^{(s,p)}}[p_{\widetilde{z}},q_{\widetilde{z}},\widetilde{L}]|^{^{2}}$ by $q_{\widetilde{z}}/p_{\widetilde{z}}$, we have for the transmissivity through a triangular air gap
\begin{equation}
\label{tris}
T_{\N}^{^{(s)}}  =   \frac{p_{z_*}q_{\widetilde{z}}}{q_{z_*}p_{\widetilde{z}}}\,
  \left|t_{\N}^{^{(s)}}\right|^{^{2}} =  \left[ \underbrace{\frac{ (q_{z_*}p_{\widetilde{z}}+ p_{z_*}q_{\widetilde{z}})^{^2}}{4\,p_{z_*}q_{z_*}p_{\widetilde{z}}\,q_{\widetilde{z}}}}_{
   \mbox{\normalsize{$\beta_{\N}^{^{(s)}}$}}}\,+ \, \underbrace{\frac{(q_{z_*}q_{\widetilde{z}} + p_{z_*}p_{\widetilde{z}})^{^2}-(q_{z_*}p_{\widetilde{z}}+ p_{z_*}q_{\widetilde{z}})^{^2}  }{4\,p_{z_*}q_{z_*}p_{\widetilde{z}}\,q_{\widetilde{z}}}}_{ \mbox{\normalsize{$\alpha_{\N}^{^{(s)}}$}}}\, \sin^{\2}(p_{\widetilde{z}} \widetilde{L}) \right]^{^{-1}},
 \end{equation}
 with
\begin{equation}
\beta_{\N}^{^{(s)}} =1+ \frac{ (q_{z_*}p_{\widetilde{z}}- p_{z_*}q_{\widetilde{z}})^{^2}}{4\,p_{z_*}q_{z_*}p_{\widetilde{z}}\,q_{\widetilde{z}}}\,\,\,\,\,\,\,
\mbox{and}\,\,\,\,\,\,\,
\alpha_{\N}^{^{(s)}} =  \frac{(q_{z_*}^{^{2}}- p_{z_*}^{^{2}}) (q_{\widetilde{z}}^{^{2}} -
 p_{\widetilde{z}}^{^{2}})}{4\,p_{z_*}q_{z_*}p_{\widetilde{z}}\,q_{\widetilde{z}}} =
 \alpha_{\B}^{^{(s)}} \,\frac{p_{z_*}q_{z_*}   }{p_{\widetilde{z}}\,q_{\widetilde{z}} }\,\,.
\end{equation}
 For TM waves, we find
\begin{equation}
\label{trip}
T_{\N}^{^{(p)}}   =  \left[ \underbrace{\frac{ (q_{z_*}p_{\widetilde{z}}+ p_{z_*}q_{\widetilde{z}})^{^2}}{4\,p_{z_*}q_{z_*}p_{\widetilde{z}}\,q_{\widetilde{z}}}}_{
   \mbox{\normalsize{$\beta_{\N}^{^{(p)}}$}}}\,+ \, \underbrace{\frac{(q_{z_*}q_{\widetilde{z}}/n^{\2} + n^{\2} p_{z_*}p_{\widetilde{z}})^{^2}-(q_{z_*}p_{\widetilde{z}}+ p_{z_*}q_{\widetilde{z}})^{^2}  }{4\,p_{z_*}q_{z_*}p_{\widetilde{z}}\,q_{\widetilde{z}}}}_{ \mbox{\normalsize{$\alpha_{\N}^{^{(p)}}$}}}\, \sin^{\2}(p_{\widetilde{z}} \widetilde{L}) \right]^{^{-1}},
 \end{equation}
with
\begin{equation}
\beta_{\N}^{^{(p)}} =\beta_{\N}^{^{(s)}} \,\,\,\,\,\,\,
\mbox{and}\,\,\,\,\,\,\,
\alpha_{\N}^{^{(p)}} =  \frac{(q_{z_*}^{^{2}}/n^{\2}- n^{\2} p_{z_*}^{^{2}}) (q_{\widetilde{z}}^{^{2}}/n^{\2} -n^{\2}
 p_{\widetilde{z}}^{^{2}})}{4\,p_{z_*}q_{z_*}p_{\widetilde{z}}\,q_{\widetilde{z}}} =
 \alpha_{\B}^{^{(p)}} \, \frac{p_{z_*}q_{z_*}   }{p_{\widetilde{z}}\,q_{\widetilde{z}} }
 \,\frac{(n^{\2}+1)p_{\widetilde{y}}^{^{2}} -n^{\2}k^{^{2}}}{ (n^{\2}+1)p_{y_*}^{^{2}} -n^{\2}k^{^{2}}}
 \,\,.
\end{equation}

\section*{\normalsize V. CONCLUSIONS}
The major difficulty in positioning two parallel surfaces sufficiently close to each other  to allow resonances phenomena or to frustrate total internal reflection is controlling their distance with accuracy and precision. To overcome this difficulty, it has been suggested  a triangular configuration\cite{te6}, see Fig.\,3. This configuration was first investigated  by using spectral lamps as light source\cite{tri} and recently by using a He-Ne laser beam\cite{te6}.

 Using the well known technique of the multiple step analysis\cite{ms1,ms2,ms3} for rectangular air gaps\cite{te7,te8}, we have proposed a new formula for the transmission through a dielectric/ar/dielectric triangular system,
 \begin{equation}
 \label{tratri}
T_{\N}^{^{(s,\,p)}}=\left\{\,\beta_{\N} + \alpha_{\N}^{^{(s,\,p)}}\,
\sin^{\2}\left[ \underbrace{k\,\left( \cos\varphi \sqrt{1-n^{^{2}}\sin^{\2}\theta_*}+n\sin \varphi \sin \theta_*\right)}_{\mbox{\normalsize $p_{\widetilde{z}}$}}\,\,\,\underbrace{h\, \sin \varphi}_{\mbox{\normalsize $\widetilde{L}$}}\right]\right\}^{^{-1}}\,\,.
\end{equation}
These formulas have to be
compared with the formulas used in literature and given in  Eq.(\ref{tb3}),
\begin{equation}
\label{trabox}
T_{\B\to \N}^{^{(s,\,p)}}=\left[\,1 +\, \alpha_{\B}^{^{(s,\,p)}}\,
\sin^{\2}\left( \underbrace{k\,\sqrt{1-n^{^2}\sin^{\2}\theta_*}}_{\mbox{\normalsize $p_{z_*}$}}\,\,\,\,\underbrace{h\,\varphi}_{\mbox{\normalsize $L_*$}} \right)\right]^{^{-1}}\,\,.
\end{equation}
As observed in the introduction, the condition $\varphi \ll 1$ is a necessary condition to guarantee, in the case of diffusion, coherence and consequently resonance phenomena and, in the case of tunneling, frustrated total internal reflection, see Fig.\,3a. In this case,
\[ p_{\widetilde{z}} \,\,\approx\,\, p_{z_{*}} + \varphi \, p_{y_{*}}\,\,.\]
In looking for significant differences  between the standard formula and the {\em new} formula proposed in this paper,  we have to amplify the $\varphi$-term in  $p_{\widetilde{z}}$. This can be done  by increasing $h$ in the sine argument. To estimate for which values of $h$  the resonances obtained from  the approximated formula given in Eq.\,(\ref{trabox}) differ from the resonances calculated  from the new formula
given in  Eq.\,(\ref{tratri}), let us explicit the resonance conditions for each case
\begin{eqnarray}
\frac{2\,\pi}{\lambda}\,\sqrt{1-n^{^{2}}\sin^{\2}\theta_*}\,\,h_{\B \to \N}^{^{(m)}}\,\,\varphi &=& m\,\pi \,\,,\nonumber
\\
\frac{2\,\pi}{\lambda}\,\sqrt{1-n^{^{2}}\sin^{\2}\theta_*}\,\,h_{\N}^{^{(m)}}\,\,\varphi &=& m\,\pi\,\left(1-\,\varphi\,\frac{n\,\sin\theta_*}{\sqrt{1-n^{^{2}}\sin^{\2}\theta_*}}\right)\,\,.
\end{eqnarray}
It is clear that the maximum shift between the resonances is obtained for
\[ m\,\varphi\,\frac{n\,\sin\theta_*}{\sqrt{1-n^{^{2}}\sin^{\2}\theta_*}} = \frac{1}{2}\,\,.\]
This implies, $m\approx 1/\varphi$ and consequently  an incident point on the first interface, $h$, of the order of $\lambda\,/\varphi^{\2}$.  Another important quantity to be estimated is the distance between two successive resonances,  $\Delta h \approx \lambda\,/\varphi$. For wavelengths of the order of  $10^{^{-7}}$m, we find
\[ \begin{array}{|c|l|c|}\hline \varphi & h\,[\mbox{m}] &\Delta h\,[\mbox{mm}\,]\\ \hline \hline
10^{^{-2}} & 10^{^{-3}} & 10^{^{-2}}\\ \hline
10^{^{-3}} & 10^{^{-1}} & 10^{^{-1}}\\ \hline
10^{^{-4}} & 10 & 1\\ \hline
\end{array}
\]
In Fig.\,4 (TE waves) and Fig.\,5 (TM waves), we plot the transmission probabilities,
(\ref{tratri}) and   (\ref{trabox}), for $\varphi=0.001$ ($n=\sqrt{2}$, $\lambda=2\pi\,10^{^{-7}}\mbox{m}$) in the case of diffusion, $\theta_*=\pi/6$. Deviations are evident  for incidence points of light of order of 10 cm.

For incidence angles very close to the critical angle
\[\theta_*=\theta_*^{^{c}}-\delta^{^{2}}\,\,\,\,\,\,\,\,\,\,(\, \delta\ll 1\,)\,\,,
\]
we find
\begin{equation}
\label{limit}
T_{\N}[\theta_*^{^{c}}-\delta^{^{2}}]\,\,\,\approx\,\,\,\left\{\begin{array}{ll}
\left[\,1+\displaystyle{\frac{\gamma_{\N}^{^{2}}-1}{4\, \gamma_{\N}}} + \gamma_{\N}\,(n^{\2}-1)\,\pi^{\2}\,(h\,\varphi/\lambda)^{^{2}}\right]^{^{-1}} & \mbox{$s$-waves}\,\,,\\
\left[\,1 + \displaystyle{\frac{\gamma_{\N}^{^{2}}-1}{4\, \gamma_{\N}}} +  \gamma_{\N}\,(n^{\2}-1)\,(\pi/n^{\2})^{^2}\,(h\,\varphi/\lambda)^{^{2}}\right]^{^{-1}} & \mbox{$p$-waves}\,\,,\end{array}
\right.
\end{equation}
where
\[ \gamma_{\N} = 1 + \frac{\varphi}{\sqrt{2\,\sqrt{n^{^{2}}-1}}\,\,\delta}\,\,.\]
 In Fig.\,6 (TE waves) and Fig.\,7 (TM waves), we compare the {\em new} results with formula (\ref{tb2}). Deviations are evident for $\delta \leq \varphi$. Nevertheless, the localization, $d$, of the incoming beam gives a constraint on this parameter,
 \[ k\,\delta^{^{2}}\approx d\,\,.\]
For a beam waist of the order of mm and a wavelength $\lambda=2\pi\,10^{^{-7}}$m, we find
\[\delta \approx 10^{^{-2}}\,\,\]
and no difference can be seen between the two formulas. This suggests that differences between the standard and the {\em new} formulas can be only seen for light diffusion and not for tunneling of light. Experiments can be prepared by an appropriate choice of $\varphi$. Observe that by increasing the angle $\varphi$,  we anticipate  the shift between the resonances (for example $\varphi=0.01 \Rightarrow h\approx \mbox{mm}$) {\em but}  we drastically increase the number of resonances ($10^{^{2}}$ in a few mm). The choice of the best $\varphi$ essentially depends on the laser wavelength.

\section*{\small \rm ACKNOWLEDGEMENTS}

The authors thank the CNPq, grant PQ/2010-2013 (S.D.L), and
FAPESP, grant 2011/08409-0 (S.A.C) for financial support.
The author also thanks the referee for his suggestions and for drawing attention to reference \cite{Anderson} on the multiple scattering approach.

\newpage

\begin{figure}
\vspace*{-3cm} \hspace*{-3.5cm}
\includegraphics[width=23cm, height=29cm, angle=0]{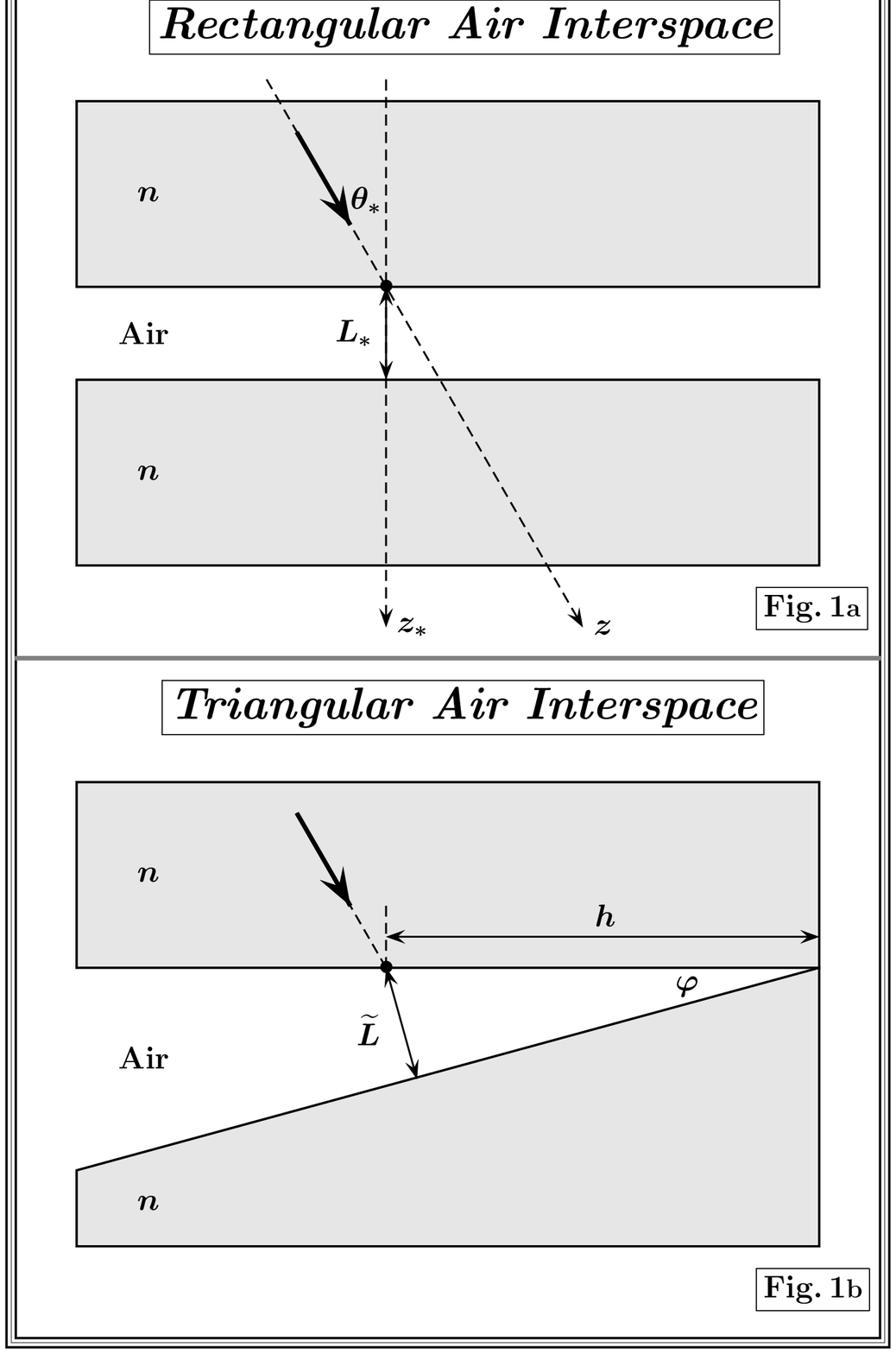}
\vspace*{-6.5cm}
 \caption{Geometric layout for the light propagation through rectangular (a) and triangular (b) air gaps situated between homogeneous media of refractive index $n$ .}
\end{figure}

\newpage

\begin{figure}
\vspace*{-3cm} \hspace*{-3.5cm}
\includegraphics[width=23cm, height=29cm, angle=0]{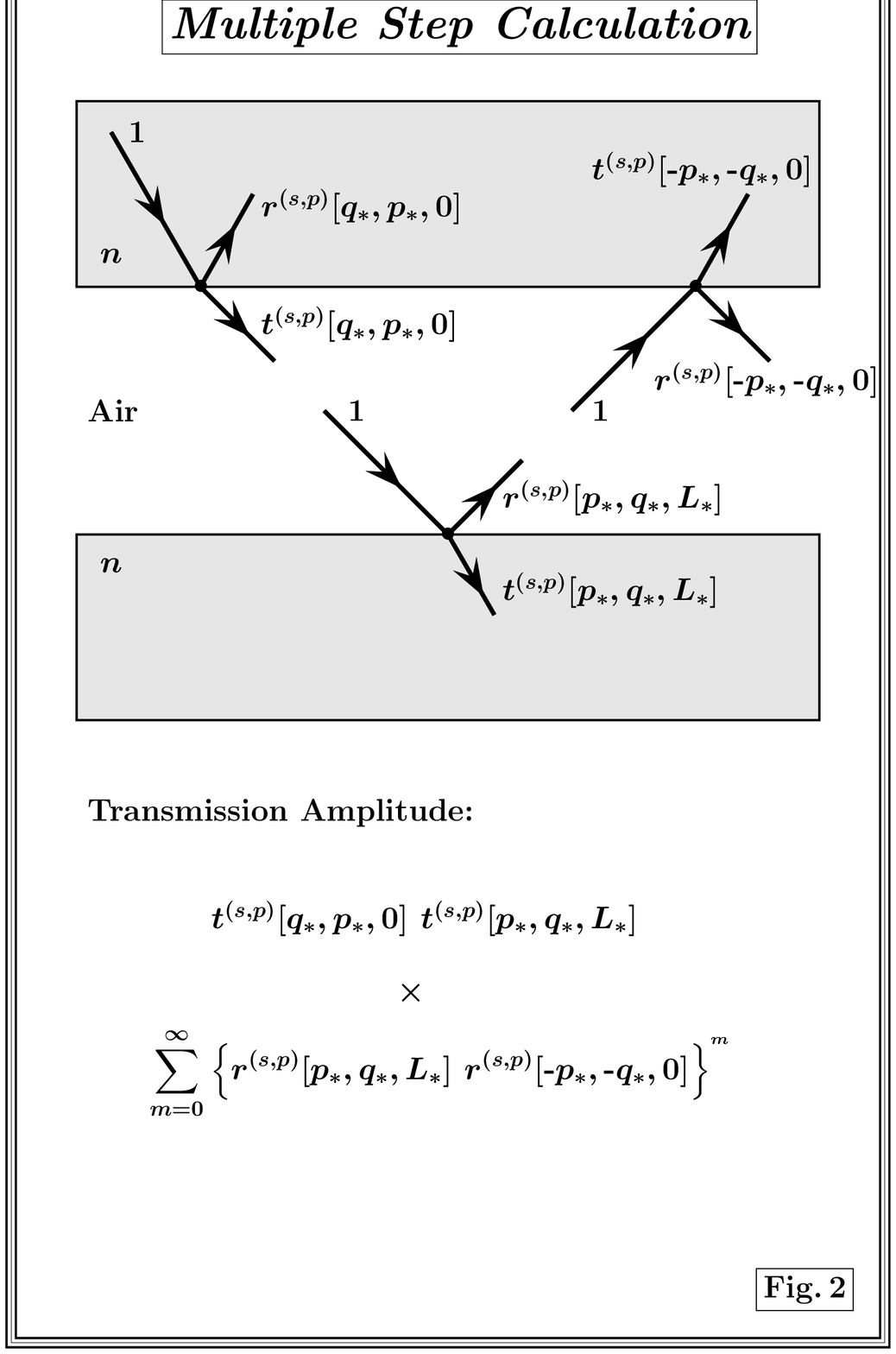}
\vspace*{-6.5cm}
 \caption{Multiple step analysis for light propagation through a rectangular air gap. This system  represents  the optical counterpart of the barrier potential problem in NRQM.}
\end{figure}

\newpage

\begin{figure}
\vspace*{-3cm} \hspace*{-3.5cm}
\includegraphics[width=23cm, height=29cm, angle=0]{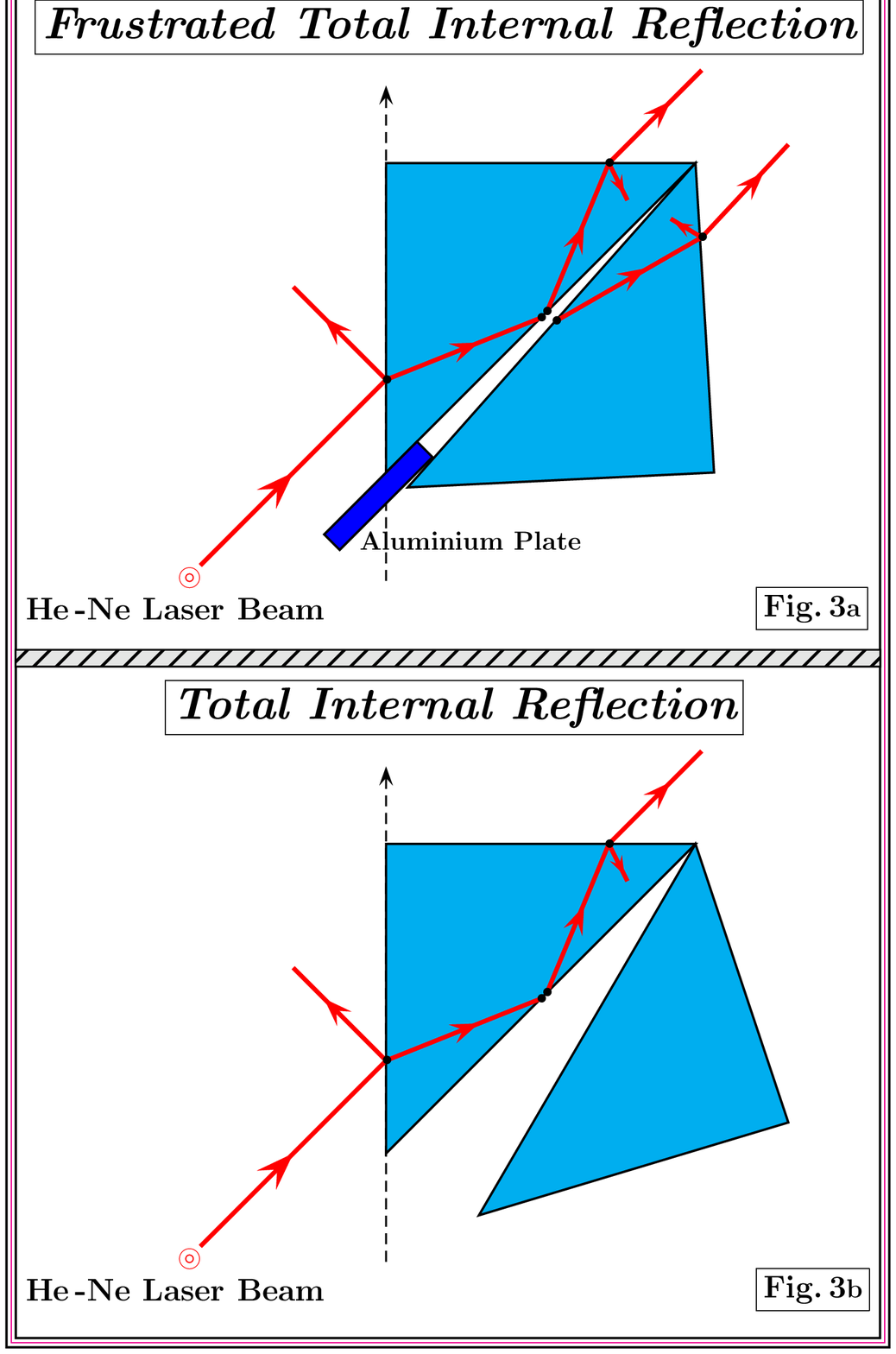}
\vspace*{-6.5cm}
 \caption{The double prism configuration recently used to analyze frustrated total internal reflection. By changing the incidence point of the light, we can study  the propagation through air gaps of different thickness and, for resonance phenomena, test our theoretical predictions.}
\end{figure}

\newpage

\begin{figure}
\vspace*{-3cm} \hspace*{-4.5cm}
\includegraphics[width=23cm, height=27cm, angle=0]{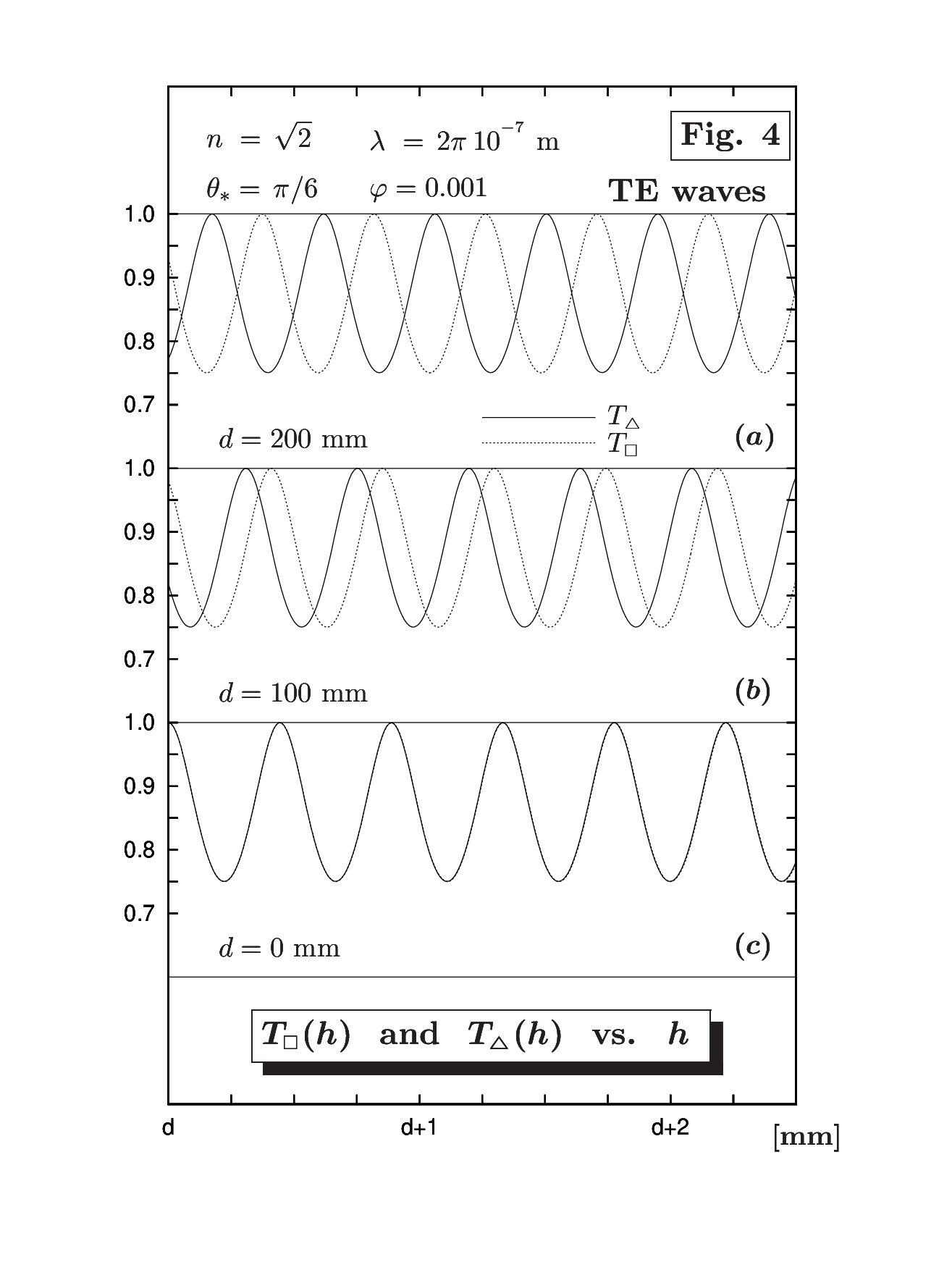}
\vspace*{-2.5cm}
 \caption{Expected resonances for the standard, $T_{\B}$, and new, $T_{\N}$, formulas for TE waves.}
\end{figure}

\newpage

\begin{figure}
\vspace*{-3cm} \hspace*{-4.5cm}
\includegraphics[width=23cm, height=27cm, angle=0]{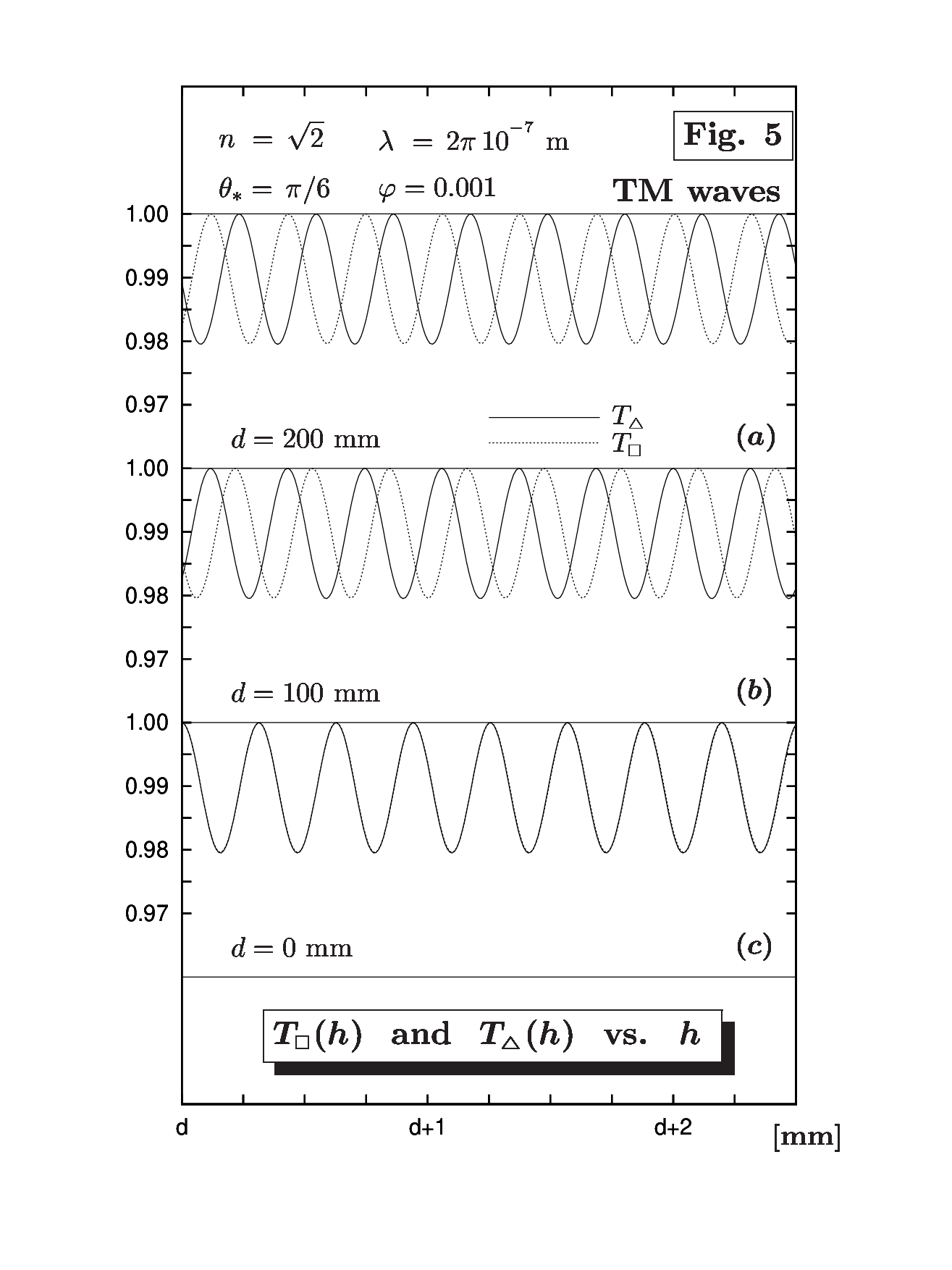}
\vspace*{-2.5cm}
 \caption{Expected resonances for the standard, $T_{\B}$, and new, $T_{\N}$, formulas for TM waves.}
\end{figure}

\newpage

\begin{figure}
\vspace*{-3cm} \hspace*{-4.5cm}
\includegraphics[width=23cm, height=27cm, angle=0]{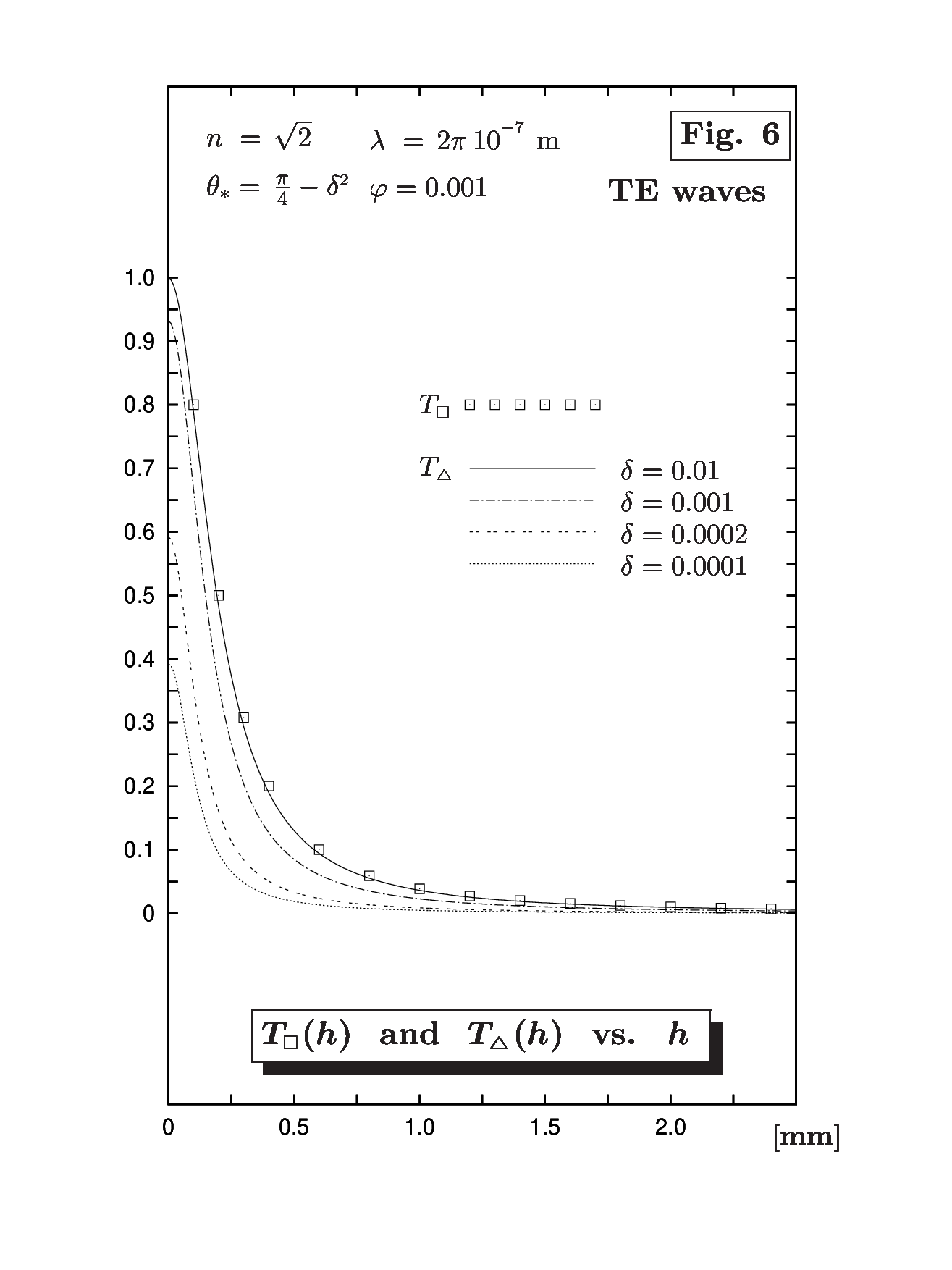}
\vspace*{-2.5cm}
 \caption{Comparison between the standard $T_{\B}$, and new, $T_{\N}$, formulas for TE waves in the case of incidence angles
 near to the critical angle $\theta_*=\pi/4$.}
\end{figure}

\newpage

\begin{figure}
\vspace*{-3cm} \hspace*{-4.5cm}
\includegraphics[width=23cm, height=27cm, angle=0]{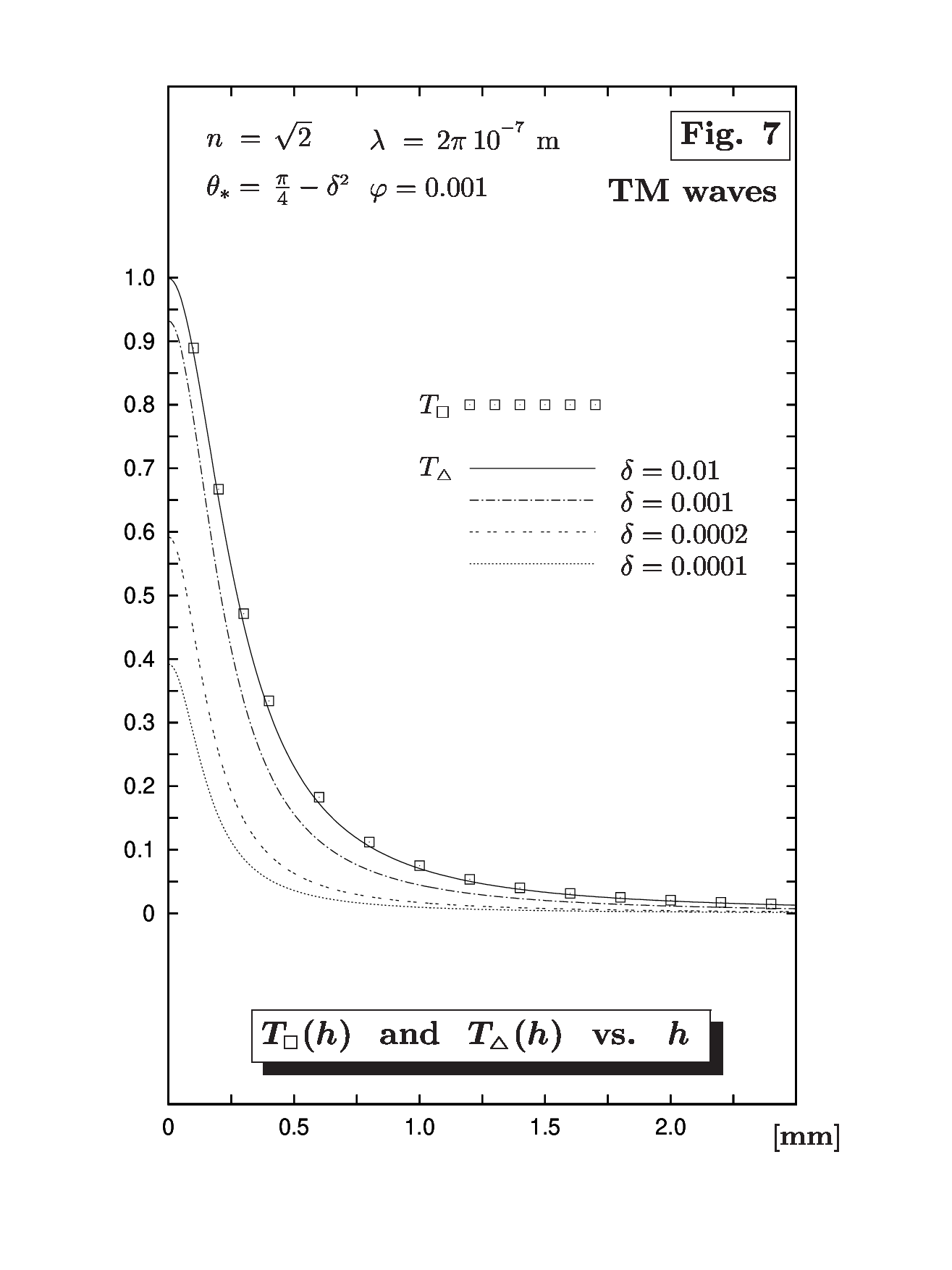}
\vspace*{-2.5cm}
 \caption{Comparison between the standard $T_{\B}$, and new, $T_{\N}$, formulas for TM waves in the case of incidence angles  near to the critical angle $\theta_*=\pi/4$.}
\end{figure}

\end{document}